\documentclass[aps,prb,amsfonts,amssymb,twocolumn,amsmath,preprintnumbers,nofootinbib,floatfix,showpacs]{revtex4}
\usepackage{amsmath,bm,amsfonts,amssymb}
\usepackage[dvips]{graphics}
\usepackage{graphicx}
\usepackage{bm}

\begin{document}

\title{Low-energy quasiparticle states at superconductor-CDW interfaces}
\author{I. V. Bobkova}
\affiliation{Institute of Solid State Physics, Chernogolovka,
Moscow reg., 142432 Russia}
\author{Yu. S. Barash}
\affiliation{Institute of Solid State Physics, Chernogolovka,
Moscow reg., 142432 Russia}

\date{\today}

\begin{abstract}
Quasiparticle bound states are found theoretically on transparent
interfaces of d-wave superconductors (dSC) with charge density wave
solids (CDW), as well as s-wave superconductors (sSC) with d-density
waves (DDW). These bound states represent a combined effect
of Andreev reflection from the superconducting side and an
unconventional quasiparticle Q-reflection from the density wave
solid. If the order parameter for a density wave state is much less
than the Fermi energy, bound states with almost zero energy take
place for an arbitrary orientation of symmetric interfaces. For
larger values of the order parameter, dispersionless zero-energy
states are found only on (110) interfaces. Two dispersive energy
branches of subgap quasiparticle states are obtained for (100) symmetric
interfaces. Andreev low-energy bound states, taking place in
junctions with CDW or DDW interlayers, result in anomalous junction
properties, in particular, the low-temperature behavior of the
Josephson critical current.
\end{abstract}
\maketitle
{\it Introduction}.
Low-energy quasiparticle states play an important role in
forming electron transport in mesoscopic hybrid superconducting
systems at low temperatures. In transparent superconductor-normal
metal-superconductor (S-N-S) junctions, subgap states originate
entirely in Andreev reflection processes. In the presence of finite
interface transparencies, both Andreev and conventional reflections
come into play in forming subgap bound states. Zero-energy Andreev
surface states in $d$-wave superconductors also represent a combined
effect of Andreev and specular quasiparticle reflections.

New interesting possibilities for forming low-energy subgap states
arise in hybrid systems involving gapped solids with various
electronic ordering like CDW or itinerant antiferromagnets (AF). In
the absence of any potential barriers and/or a Fermi velocity
mismatch, the standard specular reflection from a plane interface
vanishes. However, normal-metal quasiparticles moving with subgap
energies towards the gapped phase, will be reflected from the
transparent interface. If the gap in the quasiparticle spectrum
originates in the electronic ordering, a nonspecular quasiparticle
reflection on various plane crystal interfaces can arise in
accordance with the order parameter structure. Andreev quasiparticle
retroreflection on transparent S-N interfaces is a remarkable and
well-known effect of this kind, but it is not the only one.
Unconventional quasiparticle reflection resulting in low-energy
quasiparticle bound states arises, for example, at CDW-N interfaces
\cite{kasatkin84,visscher96,ar97}, as well as on AF-N interfaces
\cite{bhb04}.

Normal metal subgap quasiparticles change their momenta by the wave
vector $\bm Q$ of the charge density wave pattern, in an
unconventional reflection process on interfaces with the gapped CDW
phase \cite{kasatkin84}. Since the nesting condition $\varepsilon_f({
\bm k}_f+\bm Q)=-\varepsilon_f({\bm k}_f)$ is presumably satisfied in
the CDW solid, at least with the quasiclassical accuracy, the
velocity $\partial\varepsilon_f/\partial{\bm k}$ changes its
sign in a {\it $Q$-reflection} event. Then $Q$-reflection
represents a retroreflection of quasiparticles. A neutral
electron-hole pair with the trasferred momentum $\bm Q$ arises in the
$Q$-reflection event and forms the condensate in the CDW solid. No
electric current appears in this process, while in Andreev
reflection an incoming electron is reflected as a hole and a Cooper
pair carries the electric charge $2e$ into the bulk of the
superconductor. On the other hand, the retroreflection leads to a
possibility for forming quasiparticle bound states in CDW-N-CDW
systems with the same spectrum as for Andreev bound states in S-N-S
structures \cite{visscher96,ar97}. $Q$-reflection contributes also
to the conductance of N-CDW-N junctions \cite{rejaei96}. The excess
resistance in CDW-N junctions at low-voltages has been observed
experimentally and attributed to $Q$-reflection processes, when the
incident electron returns along its original path with its charge
unchanged \cite{latyshev96}. As this has been demonstrated recently
in Ref. \onlinecite{bhb04}, normal-metal quasiparticles experience a
spin-dependent $Q$-reflection from interfaces with itinerant
antiferromagnets. Quasiparticle subgap states located near interfaces
with AF have been found, in particular, near S-AF interfaces.

In the present paper we determine subgap states representing a
combined effect of Andreev and $Q$-reflections on transparent
interfaces between a semiconductor with charge density waves and a
superconductor (CDW-S). For simplicity, we consider symmetric
interfaces having identical crystal orientations on both sides. All
phases are assumed to be (quasi) two-dimensional, taking place on a
square lattice. In particular, we study below solids with a
two-dimensional CDW ordering, as well as with a $d$-density wave
phase, which has been suggested regarding the pseudogap state in
cuprates (see \cite{morr01,wang01,nayak02,trunin04,macridin04} and
references therein). We demonstrate that quasiparticle subgap
states arise on dSC-CDW and sSC-DDW (and do not appear on sSC-CDW and
dSC-DDW) interfaces. We discuss an interface between low-temperature
$s$-wave superconductor and a $d$-density wave phase implying a low
doping range in cuprates. If the order parameter for a density wave
state is much less than the Fermi energy, the quasiclassical theory
can be applied to describing the state. Within this framework, zero
energy bound states take place for an arbitrary interface
orientation. For larger values of the order parameter, the $S$-matrix
approach is developed to solving the problem in question. Then
dispersionless zero-energy states are found only on (110) interfaces.
Two dispersive energy branches of subgap quasiparticle states are
obtained for (100) interfaces. Andreev low-energy bound states,
taking place in junctions with CDW or DDW interlayers, result in
anomalous junction properties, in particular, the anomalous
low-temperature behavior of the Josephson critical current.

{\it CDW-S interfaces}.
We consider a tight-binding model for electrons with a
superconducting $\Delta^{ij}$ and a density wave $W^{ij}$ order
parameters on a square lattice
\begin{eqnarray}
H&=&-t\sum_{\langle ij\rangle\sigma} c_{i\sigma}^\dagger
c_{j\sigma} + \sum_{ i,j}(\Delta^{ij}c_{i\uparrow}^\dagger
c_{j\downarrow}^\dagger+h.c.) \nonumber \\&& +
\sum_{\langle ij\rangle\sigma} W^{ij}c_{i\sigma}^\dagger c_{j\sigma}
-\mu\sum_i c_{i\sigma}^\dagger c_{i\sigma} \enspace .
\label{H}
\end{eqnarray}
Assume nearest neighbour hopping, and consider either $s$-wave
pairing $\Delta^{ij}=-V_s\langle c_{i\downarrow}c_{i\uparrow}\rangle
\delta_{ij}=\delta_{ij}\Delta_s^{i}$ or $d$-wave pairing $\Delta^{ij}
=-V_d \langle c_{i\downarrow}c_{j\uparrow}\rangle =\Delta_d^{ij}
\delta_{|i-j|,1}$ such that $\Delta_d^{i i\pm{\hat a}}=-\Delta_d^{i
i\pm{\hat b}}$. Here $\hat a$ and $\hat b$ are basis vectors for the
square lattice with the lattice constant $a$. The order parameter for
a two-dimensional CDW is taken in the form $W^{ij}=(-1)^{i_a+i_b}
W^{i}_s\delta_{ij}=-(V^{CDW}/2)\langle n_{i\uparrow}+ n_{i\downarrow}
\rangle\delta_{ij}$, whereas for a d-density wave $W^{ij}=i(-1)^{i_a+
i_b}W_d^{ij}\delta_{|i-j|,1}=(V^{DDW}/2)\sum_{\sigma}\langle c_{i
\sigma}^\dagger c_{j\sigma} - h.c.\rangle$ and $W_d^{ii\pm\hat a}=-
W_d^{ii\pm\hat b}$. Thus, we study only the simplest model for pinned
two-dimensional charge density waves with the characteristic wave
vector ${\bm Q}=(\pi,\pi)$ on a square lattice. Although realistic
two-dimensional CDW ordering usually takes place in more complicated
situations \cite{valla04,bovet04,roca04}, the main conclusions of the
present paper can be qualitatively applicable to them as well. Thus,
if the nesting condition is valid only on a part of the Fermi
surface, just respective electrons will participate in the density
wave ordering and the Q-reflection will take place for respective
region of momentum directions. Further, the quasiclassical
superconducting $d$-wave order parameter $\Delta_d({\bm k}_f,x_i)=2
\Delta_d^{ii+\hat{a}}[ \cos (k_{fa} a)-\cos(k_{fb} a)]$ would
change its sign in a $Q$-reflection event $\Delta_d({\bm k}_f+{\bm
Q},x_i)=-\Delta_d({\bm k}_f,x_i)$ for a wide range of possible wave
vectors $\bm Q$, not only for the particular value $(\pi,\pi)$.

In describing plane interfaces, it is convenient to work in a
coordinate system where axes $x$ and $y$ are chosen
perpendicular and parallel to the interface, respectively. For a
(100) interface $x$ and $y$ coincide with the crystal axes. Then the
normal state electron band $\xi({\bf k})=-\mu - 2 t (\cos k_a +\cos
k_b)$ and the respective Brillouin zone is spanned by $k_{a,b} \in
[-\pi,\pi]$, where momenta are given in units of $a^{-1}$. For a
(110) interface we have $\xi({\bf k})=-\mu-4t\cos\bigl(k_x/\sqrt{2}
\bigr)\cos\bigl(k_y/ \sqrt{2}\bigr)$ and $k_{x}\in[-\sqrt{2}\pi,
\sqrt{2}\pi]$,\, $k_{y}\in[-\pi/\sqrt{2},\pi/\sqrt{2}]$, on account
of the periodic conditions along the surface.

A density-wave order parameter $W$ is taken to be nonzero only on one
semi-infinite half-space $x<0$, while $\Delta$ may be nonzero on the
other. For simplicity, no interface potential barrier is introduced
in the problem and we consider only identical crystal-to-interface
orientations of both half-spaces, as if they formed one and the same
square lattice. A deviation from half-filling will be assumed,
first, to be equal to zero ($\mu = 0$) or negligibly small
everywhere. This guarantees the validity of the nesting condition
$\varepsilon_f({\bm k}_f+ \bm Q )=-\varepsilon_f({\bm k}_f)$ in the
normal-metal state of both solids. We assume always that the
superconducting order parameter is much less than the Fermi energy
$\Delta\ll\varepsilon_f $, so that the quasiclassical theory of
superconductivity applies to the problem in question. If a
density-wave order parameter $W$ is sufficiently large $W\gg\Delta$,
then the $S$-matrix approach can be applied to describing the
interface Andreev bound states at CDW-S and DDW-S boundaries.
There is no need to consider the parameter $W/\varepsilon_f$ to be
small within the $S$-matrix approach. The interface states can
appear, since quasiparticles with energies below the CDW or DDW gap
do not penetrate in the bulk of solids with the density waves. At the
same time, Andreev reflection does not permit subgap quasiparticles
to enter into the bulk of the superconductor.

Quasiparticles in the superconducting halfspace can be described in
terms of standard Andreev equations for Andreev amplitudes $\tilde{
\psi}^{T}(x,{\bm k}_f)\equiv (u(x,{\bm k}_f),v(x,{\bm k}_f))$
complemented with the suitable boundary conditions at a CDW-S
interface. The difference $\bm Q$ between the outgoing $\tilde{\bm
k}_f$ and the incoming ${\bm k}_f$ momenta takes place for a
quasiparticle $Q$-reflection both from the CDW or DDW phase. The
wave vector of the density wave on the square lattice is ${\bm
Q}=(\pi,\pi)$, with respect to the crystal axes. In the
$x,y$-coordinate system, ${\bm Q}=(\pi,\pi)$ for $(100)$ interface
and ${\bm Q}=(\sqrt{2}\pi,0)$ in the $(110)$ case. So, a
quasiparticle going towards the (100) boundary of an electronically
ordered phase can change its parallel to the interface momentum
component $k_{y}$ by $Q_y=\pi$ in the $Q$-reflection process or
keep $k_{y}$ unchanged in the process of specular reflection. For
(110) boundary, by contrast, parallel to the interface component $k_{
y}$ doesn't change both in $Q$- and specular reflections. For this
reason the boundary conditions for (100) and (110) interfaces differ
from each other.

For the (110) interface the boundary conditions take the form
\begin{equation}
{\tilde\psi}^{out}(0,{k}_{y}) = \left (r^e(k_y)\frac{\displaystyle 1
+ \hat\tau_3}{\displaystyle 2}+r^h(k_y)\frac{\displaystyle 1 - \hat
\tau_3}{\displaystyle 2} \right) \tilde{\psi}^{in}(0,{k}_{y})
\label{bc_CDW_110} \enspace .
\end{equation}
Andreev amplitudes $\tilde{\psi}^{in}(x,{\bm k}_f)$ involve solutions
with $v_{f,x}<0$, in contrast with $\tilde{\psi}^{out}(x,{\bm k}_f)$.
Here $v_{f,x}=(\partial \xi(\bm k)/\partial k_x)|_{\bm k=\bm k_f}$ is
the $x$-component of the electron normal-state Fermi velocity, $\hat
\tau_\alpha$ are Pauli matrices in particle-hole space.

Reflection amplitudes for electrons $r^e$ and holes $r^h$, which
enter the quasiclassical boundary conditions, are taken for the
normal-state phase of the superconducting region. At the CDW-N
interface we find the following relation between electron and hole
amplitudes: $r^{h}_{CDW}=r^{e}_{CDW}$. This differs from the
relation, which takes place at the DDW/N interface: $r^{h}_{DDW}=
-r^{e}_{DDW}$. The latter equality is a consequence of specific
time-reversal symmetry breaking in the DDW phase \cite{morr01}.
Solving standard Andreev equations for an s-wave and a d-wave
superconductors with boundary conditions (\ref{bc_CDW_110}) on the
(110) interface, we find simple results for the subgap spectrum of
quasiparticle interface bound states. There are only zero-energy
quasiparticle bound states at CDW-dSC and DDW-sSC interfaces. At the
same time, there are no subgap states at all at CDW-sSC and DDW-dSC
(110) interfaces.

The above results can be qualitatively understood as follows. There
are no subgap states at CDW-sSC interfaces, due to the absence of
interface-induced pair-breaking processes in this case. Since at
(110) interface $Q$-reflection is quite analogous to specular one,
zero-energy bound states arise at (110) CDW-dSC interfaces for the
same reason as well-known zero-energy states at an impenetrable (110)
surface of a $d$-wave superconductor. Indeed, due to the energy gap
in the CDW solid, CDW-dSC interface is impenetrable for low-energy
quasiparticles even in the absence of any interface potential
barriers. The analogy with an impenetrable (110) surface of a
$d$-wave superconductor works, in a more complicated way, also for
the zero-energy bound states at (110) DDW-sSC interfaces. This is a
pair-breaking interface, due to a time-reversal symmetry breaking in
the DDW solid. An important role in forming subgap states on SC-DDW
interfaces plays the difference $\pi$ between phases $\Theta_{e}$ and
$\Theta_h$ of reflection amplitudes $r^{e(h)}_{DDW}$ for electrons
and holes $r^{e(h)}_{DDW}=e^{i\Theta_{ e(h)}}$. The phase difference
$\Theta_{e}-\Theta_{h}$ can be effectively ascribed to the variation
of the phase of the superconducting order parameter in a reflection
event. In order to see this, one can introduce auxiliary quantities
$\tilde{u}(x, \tilde{\bm k}_f,\varepsilon)=u(x,\tilde{
\bm k}_f,\varepsilon)e^{-i\Theta_e/2}$, $\tilde{v}(x,\tilde{\bm
k}_f, \varepsilon)=v(x,\tilde{\bm k}_f,\varepsilon)e^{-i
\Theta_h/2}$ into Andreev equations and boundary conditions, taken
for the outgoing momentum $\tilde{\bm k}_f$. Andreev amplitudes for
incoming momentum ${\bm k}_f$ are kept unchanged. Then the problem
becomes, formally, identical to the one for specularly reflecting
impenetrable boundary and the effective order parameter for the
outgoing momenta ${\Delta }_{eff}(\tilde{\bm k}_f,x)=e^{-i(\Theta_e
-\Theta_h)}\Delta(\tilde{ \bm k}_f,x)$. In general, if only a phase
difference $\Theta$ takes place between the order parameters for
incoming and outgoing quasiparticles, Andreev bound states will
appear with the energy $|\varepsilon|=|\Delta\cos(\Theta/2)|$. Since
in the problem we discuss, the $s$-wave order parameter itself does
not change its sign in a reflection process, an outgoing
quasiparticle sees an effective superconducting order parameter with
an additional phase $\Theta_e-\Theta_h=\pi$ as compared with the
phase on the incoming trajectory. This directly results in the
zero-energy interface states at (110) DDW-sSC interfaces. However, in
the case of (110) DDW-dSC interface, there is also a sign change of
the $d$-wave order parameter in a reflection event: $\Delta_d(\tilde{
\bm k}_f,x)= -\Delta_d({\bm k}_f,x)$. As a whole, an outgoing
quasiparticle sees an effective superconducting order parameter with
an extra phase $\pi -(\Theta_e-\Theta_h)$ as compared with the phase
on the incoming trajectory. Since $\Theta_e-\Theta_h=\pi$, the total
phase variation of the effective order parameter in a reflection
event vanishes. For this reason there are no pair-breaking processes
at a transparent (110) DDW-dSC interface and no interface bound
states there. We note, that the $d$-wave order parameter changes its
sign in a $Q$-reflection event for any interface-to-crystal
orientation:  $\Delta_d({\bm k}_f +\bm Q,x)=-\Delta_d({\bm k}_f,x)$.
Thus, there are no subgap states on a transparent DDW-dSC interface
with an arbitrary orientation, if only $Q$-reflection takes place
there.

Consider now (100) interfaces, for which boundary conditions can be
written as follows:
\begin{equation}
{{\tilde\psi}^{out}(0,{k}_{y}) \choose
{\tilde\psi}^{out}(0,{k}_{y}+{Q}_y)}={\check
S}{\tilde{\psi}^{in}(0,{k}_{y}) \choose
{\tilde\psi}^{in}(0,{k}_{y}+{Q}_y)} \label{bc_CDW_100} \enspace .
\end{equation}
The $S$-matrix for a CDW-N or a DDW-N boundary takes the
form $\check S = \hat S^e \frac{\displaystyle 1 + \hat
\tau_3}{\displaystyle 2}+\hat S^h \frac{\displaystyle 1 -\hat
\tau_3}{\displaystyle 2}$, where
\begin{equation}
\hat S^{e,h} = \left(
\begin{array}{cc}
r^{e,h}_{k_y,k_y} & r^{e,h}_{k_y+Q_y,k_y} \\
r^{e,h}_{k_y,k_y+Q_y} & r^{e,h}_{k_y+Q_y,k_y+Q_y} \\
\end{array}
\right) \label{S_matrix_CDW_100} \enspace .
\end{equation}

Since $Q_y\ne0$ for (100) interface, $Q$- and specular reflections
represent physically different reflection channels. Reflection
amplitudes depend now on two parallel to the surface momentum
components of incoming and outgoing quasiparticles, respectively. The
momentum components coincide with each other for specular reflection
and differ by $Q_y$ for $Q$-reflection.  One can show that $\hat S^{h
}_{CDW}=\hat S^{e}_{CDW}$ for the CDW-N interface and $\hat S^{h
}_{DDW}=\hat\rho_3\hat S^{e}_{DDW} \hat\rho_3$ for the DDW-N
interface. We define Pauli matrices $\rho_ \alpha$ in space of two
quasiparticle trajectories ($\bm k$,${\bm k}+ \bm Q$). The $S$-matrix
satisfies the unitarity condition $\check S \check S^\dagger=1$,
which follows from the conservation of the probability current for
each of independent quasiparticle solutions.  The unitarity of the
$S$-matrix leads, in particular, to the following equations: $|r^{e,h
}_{k_y,k_y}|^2=|r^{e,h}_{k_y+Q_y,k_y+Q_y }|^2=R^{sp}(k_y)$, $|r^{e,h
}_{k_y+Q_y,k_y}|^2=|r^{e,h}_{k_y,k_y+Q_y} |^2=R^Q(k_y)$. Here $R^Q$
and $R^{sp}$ are reflection coefficients for $Q$- and specular
reflections respectively. For subgap quasiparticles $R^{sp}(k_y)+R^Q
(k_y)=1$.

\begin{figure}[!tbh]
\centerline{\includegraphics[clip=true,width=3in]{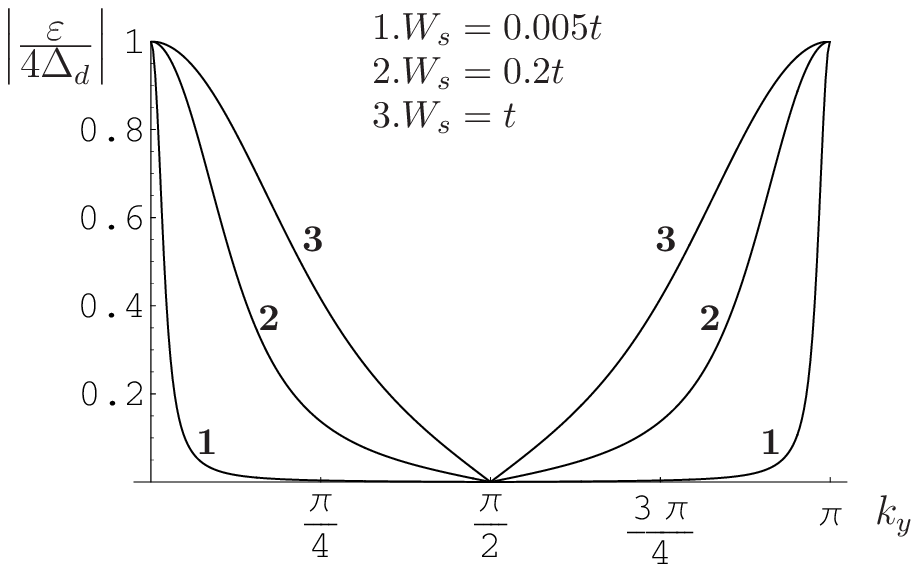}}
\caption{The dispersive bound state energy $|\varepsilon_{CDW-dSC}(
\bm k_f)|$, as a function on $k_y$ at (100) CDW-dSC interface, taken
for three values of $W_s$:\, 1.\ $W_s=0.005t$,\ 2.\ $W_s=0.2t$,\ 3.\
$W_s=t$.}
\label{CDW_bs_100}
\end{figure}

\begin{figure}[!tbh]
\centerline{\includegraphics[clip=true,width=3in]{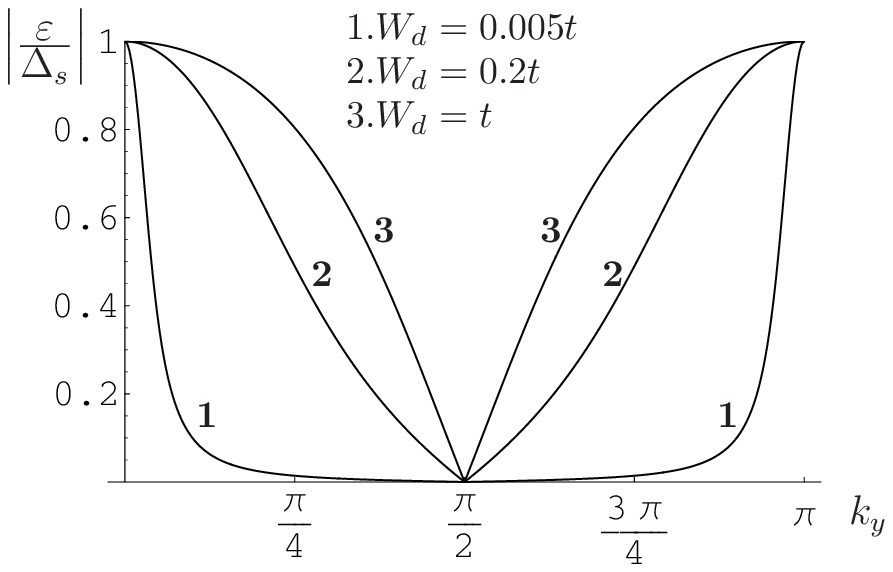}}
\caption{The dispersive bound state energy $|\varepsilon_{DDW-sSC}(
\bm k_f)|$, as a function on $k_y$ at (100) DDW-sSC interface, taken
for three values of $W_d$:\, 1.\ $W_d=0.005t$,\ 2.\ $W_d=0.2t$,\ 3.\
$W_d=t$.}
\label{DDW_bs_100}
\end{figure}

Solving Andreev equations for an s-wave and a d-wave superconductors
with boundary conditions (\ref{S_matrix_CDW_100}) on (100) interface,
we find the following results. There are no bound states at CDW-sSC
and DDW-dSC interfaces, analogously to the case of (110) interface.
However, on CDW-dSC and DDW-sSC interfaces there are two dispersive
energy branches of quasiparticle Andreev bound states, which are
symmetric with respect to the zero level:
\begin{eqnarray}
\varepsilon_{CDW-dSC}(\bm k_f)&=&\pm |\Delta_d(\bm k_f)|
\sqrt{R_{CDW-N}^{sp}(\bm k_f)}, \nonumber\\
\varepsilon_{DDW-sSC}(\bm k_f)&=&\pm |\Delta_s|\sqrt{R_{DDW-N}^{sp}(\bm k_f)}
\enspace .
\end{eqnarray}
Quantities $|\varepsilon_{CDW-dSC}(\bm k_f)|$ and $|\varepsilon_{DDW-
sSC}(\bm k_f)|$, taken for various values of $W_{d}$ and $W_{s}$,
are shown in Figs.\ref{CDW_bs_100} and \ref{DDW_bs_100} respectively,
as functions on parallel to (100) interface quasiparticle momentum
component.

We do not present explicit analytical expressions for the reflection
coefficient $R_{DDW-N}^{sp}(\bm k_f)$ and the dispersive subgap
energies $\varepsilon_{DDW-sSC}(\bm k_f)$ at (100) interfaces, since
they are too cumbersome. The explicit expression for the energy
subgap spectrum at (100) CDW-dSC interface takes the form:
\begin{eqnarray}
&&\varepsilon_{CDW-dSC}(\bm k_f)=\pm|\Delta_d(\bm k_f)|\times\nonumber\\
&&
\times\left({\frac{A(k_y)+\sqrt{A^2(k_y)+4\left(\frac{\displaystyle W_s}{
\displaystyle 2t}\right)^2}}{A(k_y)+2\sin^2 k_y
+\sqrt{ A^2(k_y)+4\left(\frac{\displaystyle W_s}{\displaystyle2t}
\right)^2}}}\right)^{1/2}
\label{bs_CDW_100} \enspace ,\nonumber\\
&&{\rm where} \nonumber\\
&&A(k_y)=\left(\frac{\displaystyle W_s}{\displaystyle 2
t}\right)^2-\sin^2 k_y \enspace .
\end{eqnarray}

As this is seen in Figs.\ref{CDW_bs_100} and \ref{DDW_bs_100}, as
well as from Eq.(\ref{bs_CDW_100}), the coefficient of specular
reflection from (100) interface in the absence of potential barriers
can become significant only if the dimensionless parameter $W_{s,d}/t
$ is of the order of unity. The appearance of specular reflection
modifies effects of $Q$-reflection and leads to a splitting of
zero-energy interface bound states. The larger the parameter
$W_{s,d}/t$, the higher the absolute value of the bound state energy.
This effect is not present at (110) interfaces, since $Q$- and
specular reflections are physically indistinguishable there, unless
an interface potential barrier and/or a Fermi velocity mismatch
result in a finite phase difference $0<|\Theta_{e}-\Theta_{h}|<\pi$.
The barrier and the mismatch open a channel of specular reflection.
This results in splitting of the zero-energy bound states at
CDW-I-dSC and DDW-I-sSC (100) interfaces. The bound state energies
reach the edge of the continuous spectrum in the limit of
impenetrable insulating interlayer. The same effect takes place at
DDW-I-sSC (110) interfaces. By contrast, bound states at (110)
CDW-dSC interfaces keep their zero energy even in the presence of any
interface potential barriers and/or a mismatch of Fermi velocities.

If parameters $W_{s,d}/t$ are sufficiently small as compared with
unity, Andreev bound states have very low energies almost in the
whole range of $k_y$, except for narrow vicinities of $k_y = 0,\pi$.
The condition $W_{s,d}/t \ll 1$ allows us to apply the quasiclassical
approach to describing the density wave phases. Similarly, the
condition $\Delta\ll t$ justifies the applicability of the
quasiclassical theory of superconductivity. Then the characteristic
lengths of the phases significantly exceed the lattice spacing
$\xi_{s,d} \equiv \hbar v_F/\Delta_{s,d}\gg a$, $\hbar v_F/W_{s,d}\gg
a$, so that the density wave amplitudes $W^{i}_s$, $W_d^{ij}$ are
also slowly varying functions as compared to the atomic scale
$a$. Below we represent a joint quasiclassical approach to the
superconducting and the density wave phases, as well as respective
results on subgap spectra.

A specific feature of the density-wave phases, which is important in
the derivation of quasiclassical equations, is associated with a
rapidly oscillating order parameter $W^{ji} \propto (-1)^{j_a+j_b}=
\exp(i{\bm Q}{\bm j})$ in the coordinate space. This prevents from
using a standard quasiclassical approach, analogously to the case of
itinerant antiferromagnets considered in this respect in Ref.
\onlinecite{bhb04}. Since $2\bm Q$ coincides with a basis vector of
the reciprocal lattice of the crystal in the absence of the density
waves, the quasiclassical equations can be written for pairs of
entangled quasiparticle trajectories $\bm k$ and ${\bm k} +{\bm Q}$.
A quasiclassical theory, modified along this way, allows arbitrary
relation between $W_{s,d}$ and $\Delta_{s,d}$, taking into account
all terms of the first order in parameters $W_{s,d}/t$, $\Delta_{s,d}
/t$. Since the gap in the energy spectrum of electrons and holes in
the CDW (DDW) phases takes place only for $|\mu|<W_s(W_d (\bm k_f))$,
we assume that the deviation from half-filling in the CDW (DDW) solid
can be finite, but not large $\mu\ll\varepsilon_f$, so that the nesting
condition $\varepsilon_f({\bm k}_f+\bm Q)=- \varepsilon_f({\bm k}_f)$
holds in the system within the quasiclassical accuracy. Then $\mu$
should be included directly in the quasiclassical equations, not in
the rapidly oscillating exponentials.

It is now convenient to  collect into a Nambu 4-spinor
the Andreev amplitudes $\psi^{T}_j\equiv (u_{j}({\bm k}_f),
u_{j}({\bm k}_f + \bm Q),v_{j}({\bm k}_f),v_{j}({\bm k}_f+ \bm
Q))$. Then the Andreev equations take the form:
\begin{eqnarray}
\left(- \mu \tau_3 \rho_0 -i\tau_3\rho_3
v_{f,x}\frac{\partial}{\partial
x} + \check W(x) + {\check \Delta}(x)\right) {\psi}(x)=\nonumber \\
= \varepsilon {\psi}(x). \label{andr}
\end{eqnarray}
Here $v_{f,x}$ is the Fermi velocity at half-filling, $\check
\Delta(x)=\check \Delta_s(x) +\check \Delta_d({\bm k}_f,x)$,
$\check\Delta_s(x)=\rho_0\Delta_s(x)
\frac{\tau_+}{2}+\rho_0\Delta^{*}_s(x)\frac{\tau_-}{2}$,
$\check\Delta_d({\bm k}_f,x)=\Delta_d({\bm k}_f,x)\rho_3
\frac{\tau_+}{2}+\Delta^{*}_d({\bm k}_f,x)\rho_3
\frac{\tau_-}{2}$. $\check W(x)=\check W_s(x) + i\check W_d({\bm
k}_f,x)$, $\check W_s(x) = W_s(x) \rho_1 \tau_3$, $\check W_d(\bm
k_f, x) = W_d(\bm k_f, x) i\rho_2 \tau_0$. DDW gap function
$W_d(\bm k_f, x) =2W_d^{ii+\hat{a}}[\cos k_{fa} -\cos k_{fb}
]$ has the same form as the $d$-wave superconducting order
parameter. Continuous coordinate $x$ in quasiclassical equations
(\ref{andr}) originated from the x-components of site positions:
$x_j=jd$. Here $d=a, a/\sqrt{2}$  for (100) and (110) interfaces
respectively.

We solve Eqs. (\ref{andr}) for superconducting and CDW regions and
match the solutions at a transparent interface at $x=0$. As one could
expect for an interface with no pair-breaking, we find no
quasiparticle interface states for a CDW-sSC interface with an
arbitrary orientation. At the same time, zero-energy bound states
arise on transparent CDW-dSC interfaces for any surface-to-crystal
orientations, since the $d$-wave superconducting order parameter
$\Delta_d({\bm k}_f)=2\Delta_d^{ii+\hat{a}}[\cos k_{fa}-\cos
k_{fb}]$ always has opposite signs for momenta $\tilde{\bm k}_{f}
={\bm k}_f+\bm Q$ and ${\bm k}_f$. This differs from a specular
reflecting impenetrable surface where a fraction of momentum
directions, for which the $d$-wave order parameter changes its sign
in a reflection event, strongly depends on a surface orientation.

The quasiclassical energy spectra exactly coincide with more general
results, obtained above for $(110)$ interfaces, and represent a good
approximation for $(100)$ interfaces under the conditions $W_{s,d}\ll
t$, $\Delta_{s,d}\ll t$. One can see in Figs.  \ref{CDW_bs_100} and
\ref{DDW_bs_100}, that even for $W_{s,d}\ll t$, $\Delta_{s,d}\ll t$
the quasiclassical approximation fails in narrow vicinities of $k_y=
0$, $\pi$. This agrees with the fact, that quasiclassical
Eqs.(\ref{andr}) do not apply in vicinities of quasiparticle momenta
where $v_{f,x}=0$. In particular, they do not apply near saddle
points of quasiparticle energies where Van Hove singularities of the
normal metal density of states take place. Since we will be
interested mostly in a transport across the interface, where the
additional factor $v_{f,x}$ arises, these momenta do not contribute
to the results noticeably and the conditions turn out not to be
restrictive.

Deviations from half-filling with $\mu\ll\varepsilon_f$
do not change the zero-energy value of the bound-state energy, within
the quasiclassical accuracy. The point is that within this framework
relations $\Theta^{CDW}_e-\Theta^{CDW}_h=0$, $\Theta^{DDW}_e-\Theta^{
DDW}_h=\pi$ are still valid for finite $\mu$. Indeed, we find,
assuming also $W\gg\Delta$, that at transparent CDW-N and DDW-N
boundaries specular reflection vanishes and there is only
$Q$-reflection for arbitrary interface orientation. The respective
quasiclassical reflection amplitudes take the form $r^e_{CDW}=r^h_{
CDW}=(\mu-i\sqrt{W_s^2-\mu^2})/W_s$,\, $r^e_{DDW}=-r^h_{DDW}=(\mu-i
\sqrt{W_d^2({\bm k}_f)-\mu^2})/iW_d({\bm k}_f)$ and satisfy required
relations.

{\it S-CDW-S tunnel junction}. Consider now Josephson junctions with
an interlayer made of gapped CDW (or a DDW) solid. Although we assume
no potential barriers in the junction, its effective transparency is
finite and tunneling of subgap quasiparticles through the gapped
phases substantially depends on the interlayer thickness $l \ll
\xi_{s,d}$. Low-energy states on the two CDW-dSC boundaries of
dSC-CDW-dSC junctions influence each other, resulting in finite
energies of interlayer quasiparticle bound states. Assuming
$W_s\gg\Delta_d({\bm k}_f)$, we find $\varepsilon_B({\bm k}_f)=\pm
\sqrt{D({\bm k}_f)}\Delta_d({\bm k}_f)\cos(\chi/2)$. Here $\chi$ is
the phase difference of superconducting order parameters on the two
banks of the junction and $D=4K/(1+K)^2$, where $K({\bm
k}_f)=\exp(-2l|W_{s}|/ |v_{f,x}|)$. In the case $\varepsilon_B \ll
\Delta_d(\bm k_f)$, the self-consistency keeps the expression for
bound states unchanged, if one introduces effective order parameters
defined in Ref. \onlinecite{barash00}. Andreev states we study in the
present paper arise as a combined effect of Andreev and
$Q$-reflections.  Contributions of these states to electric transport
are quite important. In particular, the Josephson current is entirely
carried by these states and takes the form
\begin{equation}
J=\int\limits_{-\pi/2}^{\pi/2}\frac{\displaystyle dk_y}{\displaystyle
\pi}{e\sqrt{D}|\Delta_d(\bm k_f )|\sin \frac{\displaystyle
\chi}{\displaystyle 2}}\tanh\frac{\displaystyle \sqrt{D}|\Delta_d(\bm
k_f)| \cos\frac{\displaystyle \chi}{\displaystyle 2}}{\displaystyle
2T}\enspace ,
\label{jos}
\end{equation}
which differs from the Ambegaokar-Baratoff result. In the particular
case of large interlayer width, $K,D\ll 1$, there are low-energy
states in the junction which result in low-temperature anomalous
behavior of the critical current. Under the condition $W_d({\bm k}_f)
\gg\Delta_s$, energies of quasiparticle bound states and Josephson
current in sSC-DDW-sSC junctions are obtained from the above formulas
after the substitutions $\Delta_d({\bm k}_f)\to\Delta_s$, $W_s\to W_d
({\bm k}_f)$. This behavior is similar to what can happen in tunnel
junctions with $d$-wave superconductors, S-F-S junctions with
low-energy interface states and sSC-AF-sSC junctions
\cite{bbr96,fog00,bb02,bhb04}.

{\it Acknowledgments.} We thank S.N. Artemenko for useful
discussions. The authors acknowledge the support by grant RFBR
02-02-16643 and scientific programs of Russian Ministry of Science
and Education and Russian Academy of Sciences. I.V.B. thanks the
support from the Dynasty Foundation.

\end{document}